\documentclass[aip,apl,reprint,twoside,floatfix,superscriptaddress]{revtex4-1} 

\usepackage{graphicx}
\usepackage{amsmath}
\usepackage{amssymb}
\usepackage{amsfonts}
\usepackage{dcolumn}
\usepackage{epsfig}
\usepackage{subfigure}
\usepackage{bm}
\usepackage{footnote}
\usepackage{upgreek}

\graphicspath{{device_pic/},{beam_cuts/},{measured_contours/}}

\begin{document}

\title{A Dual-polarized Broadband Planar Antenna and Channelizing Filter Bank for Millimeter Wavelengths }

\author{Roger O'Brient$^1$, Peter Ade$^{2}$, Kam Arnold$^{1}$, Jennifer Edwards$^{3}$, Greg Engargiola$^{4}$, William L. Holzapfel$^{1}$, Adrian T Lee$^{1}$, Michael J. Myers$^{1}$,Erin Quealy$^{1}$,Gabriel Rebeiz$^{3}$, Paul Richards$^{1}$, Aritoki Suzuki$^{1}$}

\address{$^{1}$Dept. of Physics, University of California, Berkeley, CA, USA.\\$^{2}$School of Physics and Astronomy, Cardiff University, Cardiff, Wales, UK.\\ $^{3}$Dept. of Electrical Engineering, UCSD, La Jolla, CA USA\\ $^{4}$ Radio Astronomy Laboratory, University of California, Berkeley, CA USA.}

\begin{abstract}
We describe the design, fabrication, and testing of a broadband log-periodic antenna coupled to multiple cryogenic bolometers. This detector architecture, optimized here for astrophysical observations, simultaneously receives two linear polarizations with two octaves of bandwidth at millimeter wavelengths. The broad bandwidth signal received by the antenna is 
divided into sub-bands with integrated in-line frequency-selective filters.  We demonstrate two such filter banks: a diplexer with two sub-bands and a log-periodic channelizer with seven contiguous sub-bands.  These detectors have receiver efficiencies of 20-40\% and percent level polarization isolation.  Superconducting transition-edge sensor bolometers detect the power in each sub-band and polarization. We demonstrate circularly symmetric beam patterns, high polarization isolation, accurately positioned bands, and high optical efficiency.  The pixel design is applicable to astronomical observations of intensity and polarization at millimeter through sub-millimeter wavelengths.  As compared with an imaging array of pixels measuring only one band, simultaneous measurements of multiple bands in each pixel has the potential to result in a higher signal-to-noise measurement while also providing spectral information.  This development facilitates compact systems with high mapping speeds for observations that require information in multiple frequency bands.
\end{abstract}

\maketitle

The steady improvement in sensitivity of millimeter and sub-millimeter wavelength bolometer arrays has enabled significant advances in the measurement of primary Cosmic Microwave Background (CMB) anisotropies, secondary CMB anisotropies such at the Sunyaev-Zel'dovich (SZ) effect, and dust emission from star-forming galaxies.  Bolometers cooled to sub-Kelvin temperatures can achieve sensitivity limited by fluctuations in the arrival rate of photons \cite{Richards}.  At this point, increasing sensitivity requires receiving more electromagnetic modes.  Over the past decade, this has been achieved using telescopes with large optical throughput (etendue) coupled to kilopixel bolometer arrays where each pixel received an independent sky signal or spatial mode.  Each detector in these arrays typically has a frequency bandwidth defined by a single band-defining filter which limits the number of spectral modes received by a telescope of fixed etendue.

However, spectral information is required by many of these imaging applications.  CMB polarization measurements require spectral information to identify and remove polarized galactic foregrounds based on spectral signatures \cite{foreground_removal}.  The thermal SZ (tSZ) effect has a null at 220 GHz and measurements in this band are particularly useful for separating the tSZ signal from primary CMB anisotropy and sources of foreground emission.  Surveys of star forming regions in high-redshift galaxies \cite{Blain} require spectral information to constrain redshift.  In principle, CII emissions redshifted into the sub-millimeter range can trace matter during reionization \cite{Gong_reionization}.  With spectral information, maps of this emission line at different redshifts could provide a history of this epoch.

In many current bolometric receivers, each pixel is sensitive to only one frequency \cite{Kuo} \cite{Niemak} \cite{myers}.  In these systems, multifrequency response can be achieved by dedicating parts of a single focal plane to different frequencies, repeating observations with different band pass filters, or even by using multiple telescopes for simultaneous observations.  Because optical design constraints limit the etendue of a given telescope, in these designs there is a necessary tradeoff between the number of spectral bands and the mapping speed of each band.  In principle, dichroic beam splitters and multiple focal planes can provide multichannel response, but such complex optical arrangements are challenging to implement while maintaining low loss and low polarization systematics.  They also tend to be large, making them poor options for upgrading existing telescopes or for proposed satellites.

In this letter, we present a fully integrated prototype pixel lithographed in thin films.  The planar antenna receives two linear polarizations over a bandwidth of two octaves. This bandwidth is partitioned with a superconducting Niobium microstrip channelizer into sub-bands which are measured with separate Transition Edge Sensor (TES) bolometric detectors.  The bolometers operate at $0.25\,$K and simultaneously measure the power in each sub-band and polarization.  We first present the design of each subcomponent in the pixel, then fabrication details of the prototype pixels, and finally the results of laboratory tests.

\begin{figure}[htbp]
\centering
\includegraphics[scale=1]{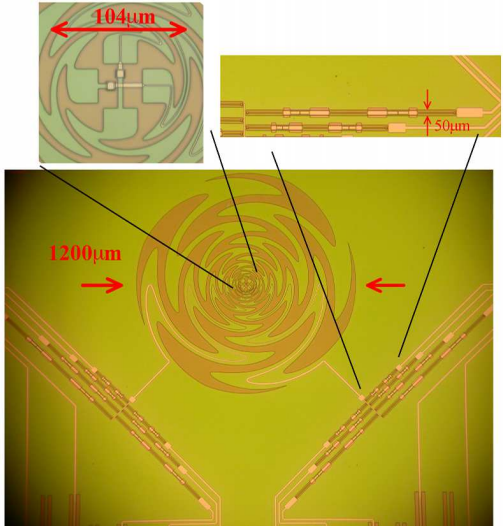}
\caption{Microscope photograph of the channelizer pixel.  Bolometers are beyond the bottom of the photo.  Upper-left: Close-up microstrip feed-points in the antenna center.  Upper-right: Close-up of one channelizer filter. \label{pixel_pic}}
\end{figure}

\begin{figure}[htbp]
\centering
\includegraphics[scale=1]{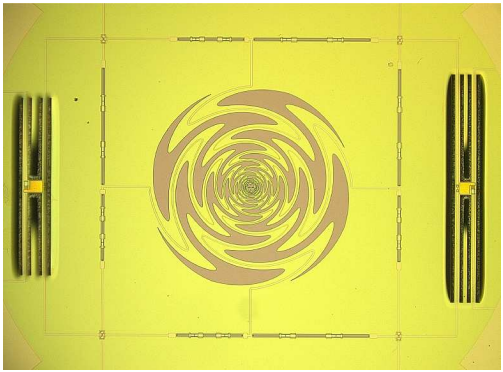}
\caption{Microscope photograph of the Diplexer pixel.  \label{pixel_pic_triplexer}}
\end{figure}

The antenna shown in Figures \ref{pixel_pic} and \ref{pixel_pic_triplexer}, is a four-armed dual-polarized sinuous antenna\cite{DuHamel} with a self similar, or log-periodic, spiral geometry that repeats every 30\% in radius. This geometry results in continuous bandwidth from 60-240 GHz, limited only by the outer and inner radii.  Our first generation devices had an outer and inner radii of 600 and $60\,\upmu\mbox{m}$.  To obtain an antenna pattern with high circularity, we find extra cells at the outside and inside of the antenna are required for the low and high frequency range respectively.   In three-dimensional electromagnetic simulations (HFSS), the extra cells reduce spurious radiating current density at the ends of the antenna, suggesting that these cells change the boundary conditions for the active cells.  For CMB polarimetry, we require percent-level ellipticities and find that we need a larger aspect ratio of inner to outer radius than was chosen for our first generation devices .  For the antenna in the diplexer device described below, we used conservative outer and inner radii of 1500 and $15\,\upmu\mbox{m}$.

We fabricated this antenna by etching slots through an otherwise continuous ground plane of 0.3 $\upmu\mbox{m}$  thick superconducting niobium on a silicon substrate.  We simulated the input impedance in Agilent's planar electromagnetics method of moments software (ADS momentum).  For a semi-infinite substrate, it lies on a complex circle of 10 $\Omega$ radius around 104 $\Omega$, which matches the impedance of the lithographed microstrip lines, leading to a return loss at the antenna feed point of less than -15 dB across the band.  This relatively constant impedance is the result of the self-complimentary geometry of the sinuous antenna in free space (no dielectrics) which is only slightly changed with the introduction to the silicon substrate \cite{DuHamel}.

We mounted a 14 mm diameter extended hemispherical silicon lens over the antenna in each device to produce beams that would match typical telescope optics and also pass through the window of our test cryostat with minimal vignetting.  We offset the hemisphere's center from the antenna center along boresight with a 2.7 mm silicon extension to approximate an ellipsoidal lens\cite{Filipovic}\cite{Edwards}. This extension was formed by the detector substrate and an additional silicon wafer.  Both the high index of the lens and the wide beamwidth of the unlensed antenna suppress the backlobe response; ADS momentum simulations suggest it is only 9\% of the total beam.  Lens-coupled detector arrays may require lenses smaller than 14mm to optimize for mapping speed, but such lens designs still provide ample room to fit the filters and bolometers under the lens in addition to the antenna.  For example, the Polarbear Experiment's f/1.8 150GHz telescope uses 6.5mm diameter contacting lenses over its detectors, which greatly exceeds the 3mm diameter of this sinuous antenna. \cite{Kam}

Each pair of opposite antenna arms couples one of two linear polarizations to microstrip transmission lines with 0.6 $\upmu\mbox{m}$ superconducting niobium (Nb) upper conductors separated from the ground plane by an insulating 0.5 $\upmu\mbox{m}$ thick silicon dioxide ($\mbox{SiO}_2$) film.  The antenna arms are defined by slots in the ground plane.  The microstrip transmission lines carry millimeter waves from the central feed points shown in Figure 1, using the back of the Nb antenna arms as their ground plane.  While traditional feeds for the sinuous antenna use transmission lines perpendicular to the antenna plane, this feed architecture makes the entire microwave design planar.  Opposite arms are driven differentially to receive only the odd mode and reflect away the even mode.

At the exterior of the antenna, each polarization's transmission lines enter a filter bank that partitions the bandwidth into narrow channels.  For terrestrial experiments which must observe through the Earth's atmosphere, each band must be located in a spectral window of high transmittance to avoid undesirable atmospheric emission and noise\cite{OBrientLTD12}. In one such design, we extract bands at 90 and 150 GHz using a diplexer circuit that splits the line in a microstrip T-junction leading to two three-pole filters.  We have also fabricated and tested a three-band triplexer version (90, 150, and 220 GHz), with similar band positions and efficiencies.  The inductors in these filters are short stretches of high impedance transmission line separated by pi-networks of capacitors.  This architecture allowed us to form the capacitors from the same Nb and $\mbox{SiO}_2$ films used for the microstrip lines in a way that does not require vias between the Nb films.  We terminate the filtered signal from a differential pair of transmission lines by differentially driving a 20 $\Omega$ lumped resistive load in close thermal contact with a TES on a thermally isolated bolometer island on the silicon wafer.  Both the resistive load and TES are etched from the same aluminum-titanium bilayer with a superconducting transition between 0.5 and 0.6 K.

In an alternative design for satellite and balloon applications where atmospheric loading is less of a concern, each polarization's transmission line enters a channelizer with three-pole band-pass filters of geometrically decreasing resonant frequencies \cite{Galbraith_3pol}, where the individual resonant filter design is nearly identical to that used in the diplexers and triplexers described above.  We connect these filters by a trunkline of series inductors formed from short stretches of 15 $\Omega$ microstrip.  For frequencies that resonate in filters in the interior of the manifold, the filters near the entrance port will appear capacitive.  This shunt capacitance acts with the series trunk-line inductors to create an effective transmission line circuit that passes the mode along until it reaches its resonant filter. The log-periodic nature of the channelizer guaranties a low return loss over the seven bands, provided that the input port is tuned with a shunt capacitor.  We designed this channelizer to function as a low-resolution spectrometer with low cross-talk between channels;  we could alternatively allow increased spectral cross-talk between tighter packed channels in a design more suitable for spectroscopy\cite{Galbraith_1pol}.  We terminated each filter with a lossy aluminum-titanium transmission line on the released bolometer, also formed from the same bilayer as the TES. In all of our designs, we voltage-bias the TES bolometers into the superconducting transition and measure current changes with a SQUID ammeter\cite{Irwin}.  Strong electrothermal feedback holds the TES in the transition, elevated above the thermal bath temperature of 0.3 K.  While we chose to use TES bolometers because of their proven low noise performance, in principle we could replace the bolometers with other detectors, such as kinetic inductance detectors\cite{Zmuidzinas_KIDS_review}.

\begin{figure}[htbp]
\centering
\includegraphics[scale=1]{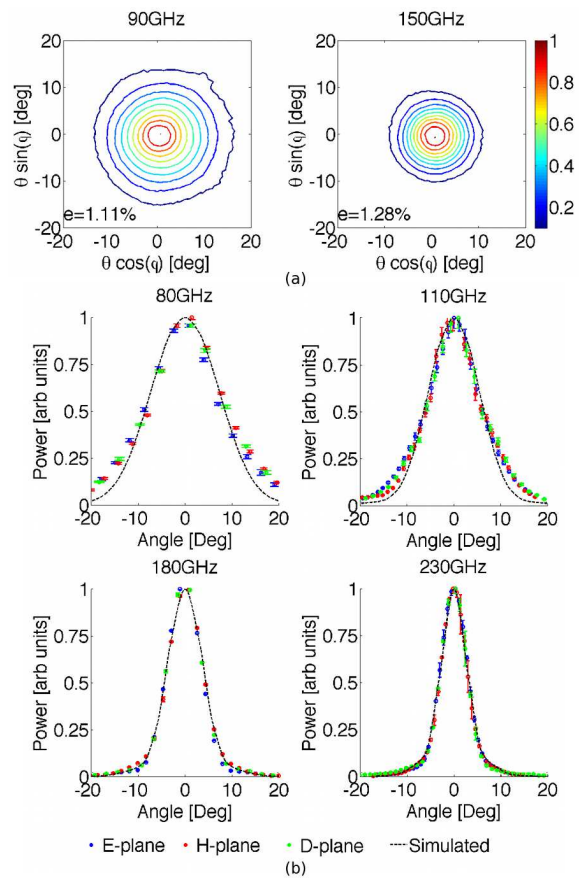}
\caption{Far-field antenna patterns averaged over the 30\% bandwidth integrated filters. (a) Measured 2-D Diplexer maps through a diplexer, stamped with ellipticity. (b) Measured beam profiles compared to simulations at four frequency channels show acceptable beams over the entire antenna operating band.  The simulations average the beams over the filter bandwidth.\label{all_beams}}
\end{figure}

Our test cryostat has a foam window and a stack of metal-mesh filters to reduce IR loading.  We scanned a thermal source modulated between 77 and 300 K  on a pair of translation stages in front of the cryostat to map the far-field beam patterns from each channel in our test pixels.  The 
first device we characterized was designed to detect power in each polarization through a diplexer with bands centered  at 90 and 150 GHz. Two-dimensional maps from one polarization in each band are shown in Figure \ref{all_beams}.  The other devices that we mapped contained only a single band-pass filters between the antenna and bolometer so we could demonstrate acceptable antenna performance up to the 250 GHz upper band edge.  Figure \ref{all_beams} shows measured and simulated single-plane beam cuts through those single filter devices.  All beams were highly circular with an ellipticity (difference divided by sum of major and minor axis widths of the best fit Gaussian) of only $\sim 1$ percent.

We find good agreement between beam measurements and simulations, which we co-plotted in Figure \ref{all_beams}.  We simulated the antenna on a semi-infinite silicon substrate, and then accounted for the lens surface with an optics software that modeled refraction and diffraction at the vacuum-silicon surface \cite{Filipovic}.  The software ignores reflections which, for a synthesized ellipse, only slightly alter the beam and are greatly suppressed by an anti-reflection coating.   The beam opening angle, or antenna gain, varies with wavelength since the lens acts as a diffracting aperture with a constant physical size.  In the shorter-wavelength channels, the lens is an electrically larger diffracting surface, which narrows the beam relative to the longer-wavelength channels\cite{Edwards}.  When used in a focal plane array where the highest frequency beam matches the telescope optics, the lower frequencies will have a relatively large beam and lower spillover efficiency. But provided that the aperture stop is cold, this loss is offset by the higher pixel density of those channels relative to the pixel density of monochromatic arrays.

We positioned the chopped source on boresight and rotated a wire-grid polarizer between the window and source aperture, searching for the minimum power position in each channel.  This test showed polarization isolation of 2.8\% or lower, and that highest value occurred in the higher frequency channels.  Based on our experience with adding exterior cells to improve the lower band edge performance, we expect that this leakage will decrease with the addition of more high-frequency cells.  The diplexer device showed 0.3\% and 1.6\% polarization isolation in the 90 and 150 GHz channels.

\begin{figure}[htbp]
\centering
\includegraphics[scale=1]{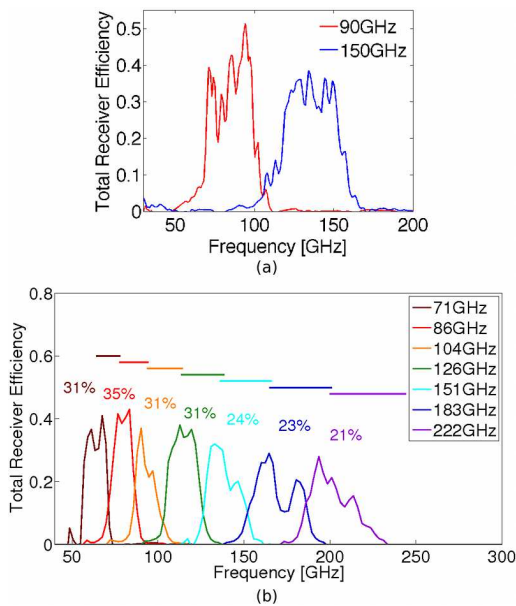}
\caption{Measured spectra of each channel.  (a) shows two channels from one polarization of a diplexer device.  (b) shows one polarizations of a log-periodic channelizer.  We show the designed -3 dB bandwidths in the horizontal bars above the plots and have printed the band-averaged efficiency above each.  We measured the Receiver Efficiency using the power received from a temperature modulated beam-filling thermal load and used these to normalize the spectra\label{spectra}}
\end{figure}

Finally, we measured the spectral response of all the channels in our channelizer with a Fourier transform spectrometer that filled the antenna beam for all but the lowest frequency channel.  Figure \ref{spectra} shows the results of this spectroscopy for both the diplexer and log-periodic channelizer, where we have divided away the spectral response of the interferometer's 0.01 inch thick mylar beam-splitter.

\begin{table}[htbp]
\caption{Sources of loss in Diplexer Measurement}
\centering
\begin{tabular}{c c c}
\hline\hline
$Component$&\multicolumn{2}{c}{$Efficiency$} \\[0.5ex]
$$ & 90GHz& 150GHz\\ [0.5ex]
\hline
Cryostat Thermal Filters& $\approx$70\%& $\approx$70\%\\
Lens-vacuum interface & 90\%&  90\%\\
Antenna Front-lobe & 91\%& 91\%\\
Dielectric Loss ($\tan(\delta)$=0.005)& 81\%& 70\%\\
\textbf{Product} & 45\%& 39\%\\[1ex]
\hline
\end{tabular}
\label{Loss_LP}
\end{table}

We normalize the peak of each spectrum to the receiver efficiency measured with a chopped 77--300 K beam-filling thermal source and label each curve in figure \ref{spectra} with the band averaged efficiency.  Table \ref{Loss_LP} summarizes the losses that explain the observed efficiencies of 45\% in the diplexer.  The AR-coating was optimized for 120 GHz, which reduced the efficiency by about 5\% in both 90 GHz and 150 GHz.  Internal reflections within the coating further reduce the efficiency by another 5\% \cite{vandervost}.  The microstrip dielectric losses are a function of frequency with an average loss of 20\% between the diplexed channels.  The log-periodic channelizer had lower efficiency because we did not anti-reflection coat the contacting lens and there was an impedance mismatch between the microstrip and antenna.  We will correct these in future versions as we did with the diplexer device.

We envision using focal planes of these multi-colored pixels in CMB and sub-millimeter telescopes in lieu of focal planes with single color pixels.  Focal planes of these multichroic pixels can provide the same experimental sensitivity as several monochromatic focal planes, enabling a significant increase in sensitivity of both ground- and space-based instruments.

We fabricated these devices in the Berkeley Microlab.  NASA grant NNG06GJ08G supported this work.

\end{document}